\documentclass{article}
\usepackage{spconf, epsfig}
\usepackage{mathtools}
\usepackage{physics}
\usepackage{amssymb,amsmath,amsfonts,amsthm}

\title{QUANTUM-INSPIRED TENSOR NETWORK FOR EARTH SCIENCE}
\name{Soronzonbold Otgonbaatar, Dieter Kranzlmüller}
\address{German Aerospace Center, Ludwig-Maximilians-Universität Munich}
\begin{document}
%
\maketitle
\begin{abstract}
Deep Learning (DL) is one of many successful methodologies to extract informative patterns and insights from ever increasing noisy large-scale datasets (in our case, satellite images). However, DL models consist of a few thousand to millions of training parameters, and these training parameters require tremendous amount of electrical power for extracting informative patterns from noisy large-scale datasets (e.g., computationally expensive). Hence, we employ a quantum-inspired tensor network for compressing trainable parameters of physics-informed neural networks (PINNs) in Earth science. PINNs are DL models penalized by enforcing the law of physics; in particular, the law of physics is embedded in DL models. In addition, we apply tensor decomposition to HyperSpectral Images (HSIs) to improve their spectral resolution. A quantum-inspired tensor network is also the native formulation to efficiently represent and train quantum machine learning models on big datasets on GPU tensor cores. Furthermore, the key contribution of this paper is twofold: (I) we reduced a number of trainable parameters of PINNs by using a quantum-inspired tensor network, and (II) we improved the spectral resolution of remotely-sensed images by employing tensor decomposition. As a benchmark PDE, we solved Burger's equation. As practical satellite data, we employed HSIs of Indian Pine, USA and of Pavia University, Italy.
\end{abstract}
\begin{keywords}
Tensor decomposition, quantum-inspired tensor decomposition, quantum-inspired machine learning.
\end{keywords}
\section{introduction}
\label{sec:intro}
 
Deep Learning (DL) is a machinery for extracting most informative patterns, insights from large-scale data, and apply this knowledge to make predictions \cite{lecun}. DL models currently have been outperforming conventional techniques and methods in science and engineering, even in remote sensing and Earth science \cite{Choudhary2022, chen2020, datcu0}. However, DL models compose of a huge number of parameters, making their interpretation and predictions on large-scale data difficult. Their energy requirements also extremely limit their scalability  (or computationally expensive) \cite{patterson2021}. Hence, the authors of the articles \cite{Cichocki14,gao2020mpo, anima2022} utilized a quantum-inspired tensor network to compress the parameters (e.g., hidden layers) of DL models and to decompose data tensors in very small factor matrices. Here, tensors are multidimensional arrays which can generalize vectors and matrices. A quantum-inspired tensor network can compress the training parameters of DL models and decompose data tensors in a small number of factor matrices. It is also widely used to represent quantum Machine Learning models as tensor-networks, which can be efficiently trained on big real-world datasets on GPU tensor cores \cite{huang2022}. 

Physics-Informed Neural Networks (PINNs) are DL models (e.g., Neural Networks), whose training parameters are penalized by enforcing the law of physics \cite{RAISSI2019686}; namely, the law of physics is embedded in Neural Networks (NNs). Moreover, PINNs can be utilized to compute and analyse computationally expensive Partial Differential Equations (PDEs) when data is of limited quantity and quality \cite{Karniadakis2021}. However, PINNs are still computationally expensive for obtaining solutions to PDEs in Earth science.

Remotely-sensed datasets are data tensors $\mathcal{X}\in \mathbb{R}^{I_1\times \dots\times I_n}$ which are so complex and diverse that they cannot be easily classified and analyzed even by using DL models. In particular, these datasets are characterized by not only volume but also another so-called ``4V'' features (Volume, Variety, Veracity, and Velocity) \cite{Reichstein2019}.

\begin{figure}[t!]
\begin{minipage}[b]{1.0\linewidth}
  \centering
 \centerline{\epsfig{figure=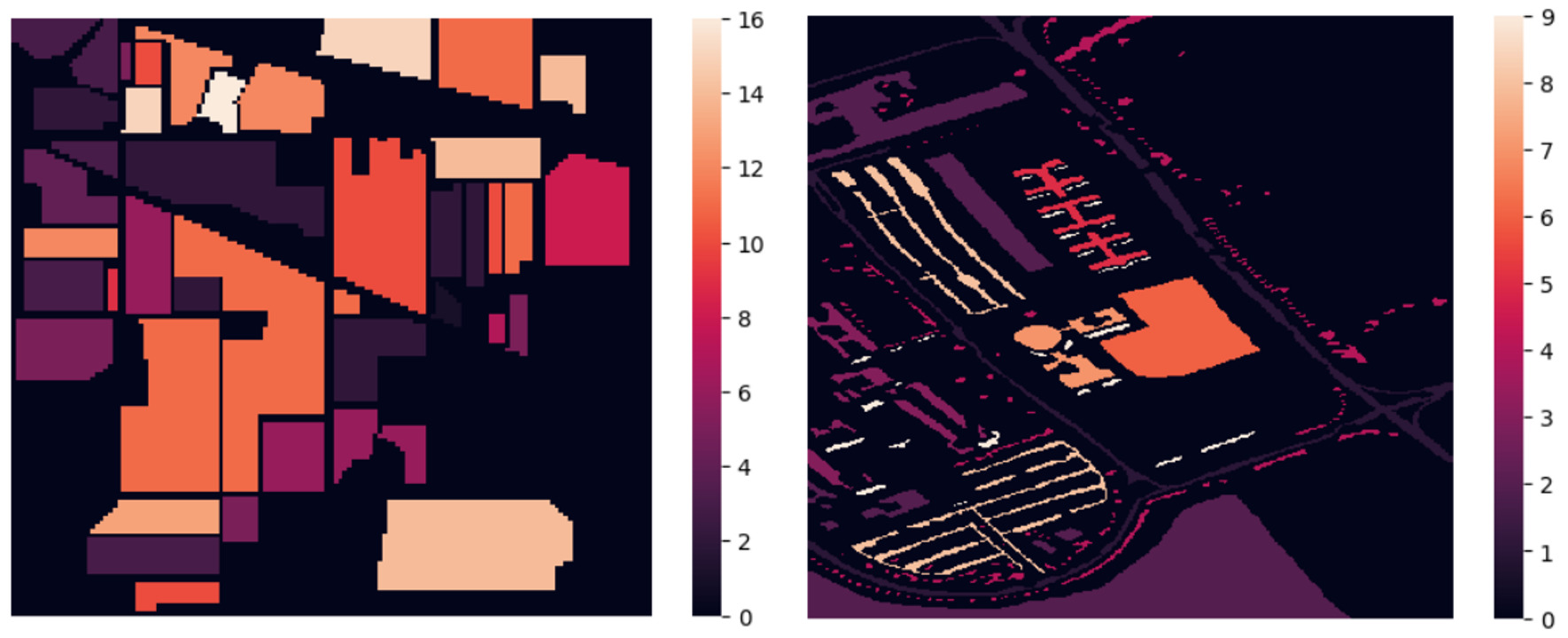,width=8.5cm}}
  \vspace{0.1cm}
  \centerline{}\medskip
\end{minipage}
\caption{Satellite datasets: [Left] HSIs of Indian Pine, USA and [Right] of Pavia University, Italy}
\label{fig: hsi}
\end{figure}

The key contribution of this paper is twofold: The first contribution of this paper is that we reduced a number of trainable parameters of DL models (i.e. PINNs) by using the quantum-inspired tensor network. The compressed DL models can be also applied to analyse and classify big real-world datasets as shown in the article \cite{gao2020mpo}. The second contribution of this paper is that we improved the spectral resolution of remotely-sensed images by employing tensor decomposition. As practical satellite data, we employed HSIs of Indian Pine, USA and of Pavia University, Italy. As a PDE, we considered Burger's equation.

%
%

\section{our datasets}

We use practical satellite datasets and refer them as 3rd-order data tensors. In particular, the HSI of Indian Pine is the data tensor $\mathbb{R}^{240\times240\times 200}$ with 16 classes, and the HSI of Pavia University is the data tensor $\mathbb{R}^{610\times340\times 103}$ with 9 classes (see Fig. \ref{fig: hsi}). 

\begin{figure}[t!]
\begin{minipage}[b]{1.0\linewidth}
  \centering
 \centerline{\epsfig{figure=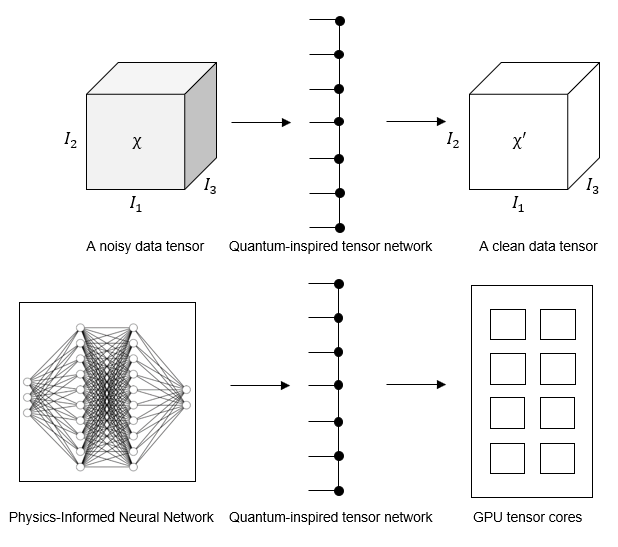,width=8.5cm}}
  \vspace{0.1cm}
  \centerline{}\medskip
\end{minipage}
\caption{The two contributions of this paper in pictorial representation: [Top] Quantum-inspired tensor network (decomposition) for improving spectral resolution of real-world noisy data tensor, and [Bottom] Quantum-inspired tensor network for compressing Physics-Informed Neural Networks, which can be efficiently simulated on GPU tensor cores.}
\label{fig: pictIII}
\end{figure} 

\section{our methodology}

Remotely sensed images can be viewed as 3rd-order data tensors $\mathcal{X}\in\mathbb{R}^{I_1\times I_2\times I_3}$. The 3rd-order data tensors can be decomposed in factor matrices by using so-called CANDECOMP/PARAFAC (CP)-decomposition \cite{Cichocki14}:

\begin{equation}\label{eq: cp}
        \mathcal{X}=\sum_{r=1}^R \mathbf{a}_r\circ\mathbf{b}_r\circ\mathbf{c}_r,
\end{equation}
where $R$, called the rank, is a real positive number, ``$\circ$'' denotes an outer product, and $\mathbf{a}_r\in\mathbb{R}^{I_1}$, $\mathbf{b}_r\in\mathbb{R}^{I_2}$, and $\mathbf{c}_r\in\mathbb{R}^{I_3}$ are factor matrices (see Fig. \ref{fig: pictIII} [Top]).

Another commonly used quantum-inspired tensor network is Tensor Train (TT)-decomposition, called also Matrix Product State (MPS) in quantum physics \cite{roberts2019}. TT-decomposition expresses a 3rd-order tensor as core tensors and factor matrices:

\begin{equation}\label{eq: tt}
        \mathcal{X}= \mathbf{A}\times_3^1\mathbf{G}^{(2)}\times_3^1\mathbf{B},
\end{equation}
where $\mathbf{G}^{(2)}\in\mathbb{R}^{R_{1}\times I_2\times R_2}$ is a core tensor, $\mathbf{A}$ and $\mathbf{B}$ are factor matrices, and $\times_3^1$ is called a mode-(k,l) product . 

TT-decomposition can compress DL models, and the compressed DL models can generate classes with the similar accuracy as their non-compressed ones \cite{gao2020mpo} (see Fig. \ref{fig: pictIII} [Bottom]). In addition, TT-decomposition is widely employed to efficiently simulate quantum circuits on conventional computers. Hence, TT-decomposition have been applying to design and train quantum-inspired machine learning models on large-scale datasets on GPU tensor cores \cite{Stoudenmire2018, Huggins2019, glasser2020}.

%
%

%
%

\begin{figure}[t!]
\begin{minipage}[b]{1.0\linewidth}
  \centering
 \centerline{\epsfig{figure=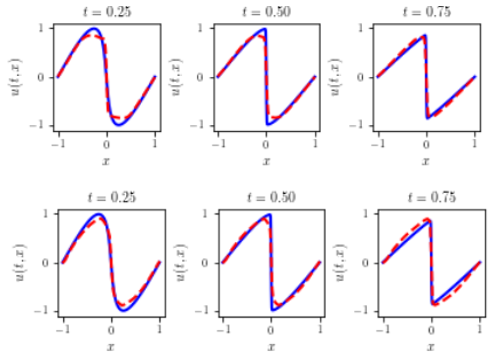,width=8.5cm}}
  \vspace{0.1cm}
  \centerline{}\medskip
\end{minipage}
\caption{A solution to Burger's equation (blue is an exact solution, and red is a predicted solution): [Top] The original PINN, and [Bottom] The compressed PINN}
\label{fig: pictIV}
\end{figure}

\section{our experiment}

\subsection{Contribution I: compressing PINNs}

We represented a solution $u=u(t,x)$ to 1D Burger's equation by an NN \cite{RAISSI2019686}. In mathematical form, 1D Burger's equation is
\begin{equation}
    \begin{split}
    &u_t+uu_{x}-(0.01/\pi)u_{xx}=0, \quad t\in[0,1],\\
    &u(0,x)=-\sin(\pi x),\\
    &u(t,-1)=u(t,1)=0.
\end{split}
\end{equation}
When we used the NN with $8$ hidden layers, and each layer comprises $100$ neurons, its trainable parameters are amounted to $71,101$ parameters.
We reduced these $71,101$ parameters to $32,701$ parameters by compressing the odd numbers of the hidden layers by utilizing the TT-decomposition (see Fig. \ref{fig: pictIII} [Bottom]) \cite{gao2020mpo}. We found a solution $u$ to the Burger's equation while utilizing both the original and compressed PINNs. Furthermore, the compressed PINN generated a solution to the Burger's equation with high accuracy such as having been produced by its original PINN, while it occupies a smaller parameter space than its original one (see Fig. \ref{fig: pictIV}). More importantly, the compressed NNs can be also utilized to analyse and classify any real-world datasets as shown in the article \cite{gao2020mpo, gao2022}.

\begin{figure}[t!]
\begin{minipage}[b]{1.0\linewidth}
  \centering
 \centerline{\epsfig{figure=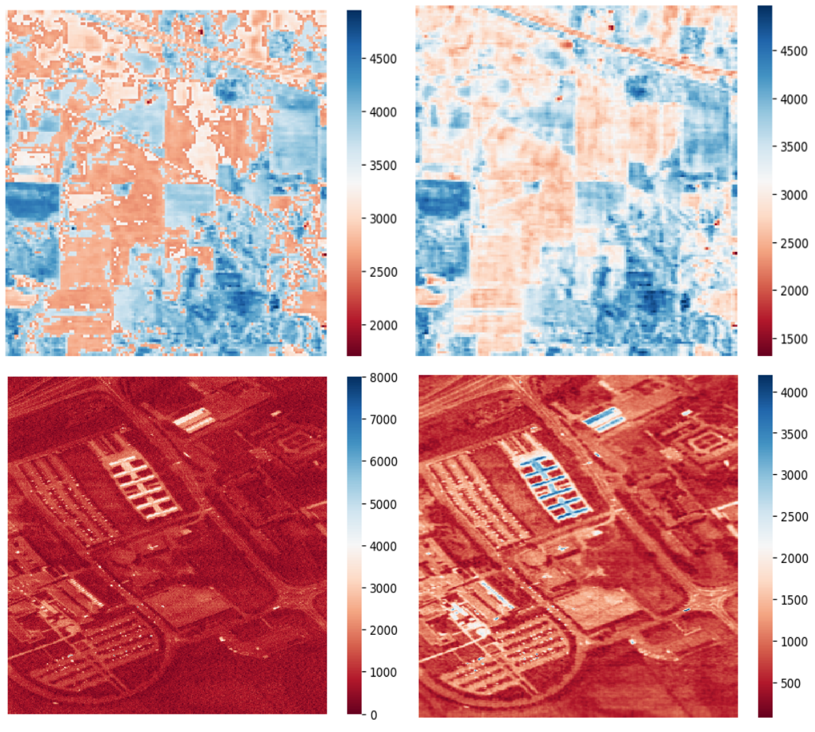,width=8.5cm}}
  \vspace{0.1cm}
  \centerline{}\medskip
\end{minipage}
\caption{Two examples of HSIs before and after tensor decomposition: [Top] The band 74 of the Indian Pine HSI before and after tensor decomposition, and [Bottom] The band 1 of the Pavia University HSI before and after tensor decomposition.}
\label{fig: pictII}
\end{figure} 

\subsection{Contribution II: decomposing real-world data tensors in factor matrices}
We decompose two practical HSIs shown in Fig. \ref{fig: hsi} in a very small number of factor matrices by using CP-decomposition expressed by Eq. \eqref{eq: cp} to improve their spectral resolution; we illustrate our method for decomposing these HSIs in Fig. \ref{fig: pictIII} [Top]. In our experiment, we set the rank $R$ of the CP-decomposition at 145.
For the Indian Pine HSI, the decomposition time was $0.1711$ seconds, the compression ratio was $60$, and the R-squared value between the raw and the decomposed Indian Pine HSI was $0.9959$.    
For the Pavia University HSI, the decomposition time was $1.1013$ seconds, the compression ratio was $140$, and the R-squared value between the raw and the decomposed Pavia University HSI was $0.9450$. From these results, we gained the insight that we improved the spectral resolution of the HSIs, and the HSIs can be stored efficiently in conventional storage devices at the same time, while applying tensor decomposition to the practical HSIs. We presented some visual examples of our finding in Fig. \ref{fig: pictII}.

\section{conclusion}
This paper focused on designing and applying a quantum-inspired tensor-network to DL models and real-world data tensors. Our contribution is twofold: (I) We reduced the parameters of a DL model when compressing them by using TT-decomposition. As a DL model, we utilized a physics-informed neural network for finding a solution to 1D Burger's equation. The compressed model generates solutions to 1D Burger's equation with high accuracy such as having produced by its original one.
(II) We improved the spectral resolution of hyperspectral images (i.e. data tensors) by decomposing them in sparse factor matrices through CP-decomposition. The decomposed data tensors are represented by sparse tensors, while the decomposition time was extremely small (around 1 second). Additionally, we can store these decomposed images (i.e. sparse tensors) efficiently and securely in distributed storage devices thanks to their sparse factor matrices. As practical HSIs, we used HSIs of Indian Pine, USA and of Pavia University, Italy.

As a future and on-going work, we invent and design quantum-inspired machine learning models for data-driven and model-driven practical problems. In addition, we invent and analyse DL models supported by quantum tensor networks \cite{Stoudenmire2018, Huggins2019, Karniadakis2021, glasser2020}.



\bibliographystyle{IEEEbib}
\bibliography{tensor.bib}

\end{document}